\documentclass[11pt]{article}
\usepackage[letterpaper,margin=1in]{geometry}
\usepackage[T1]{fontenc}
\usepackage{lmodern}
\usepackage{amsmath,amssymb}
\usepackage{graphicx}
\usepackage{booktabs}
\usepackage{array}
\usepackage[round,authoryear]{natbib}
\usepackage{caption}
\usepackage{xcolor}
\usepackage[hyphens]{url}
\usepackage[colorlinks=true,linkcolor=blue!50!black,citecolor=blue!50!black,urlcolor=blue!50!black]{hyperref}

\newcommand{\deff}{\mathrm{deff}}
\newcommand{\pp}{\,pp}

\title{How Many Sources, Not Just How Many Problems:\\
Planning the Precision of LLM Evaluations from Earlier Results}
\author{Guang Qiu\\
Department of Statistics, Florida State University\\
\texttt{gq21@fsu.edu}}
\date{}

\begin{document}

\maketitle

\begin{abstract}
When benchmark problems share sources, such as variants of one template or tasks from one repository, an evaluation
budget must fix the number of sources as well as the number of problems. We ask whether pilot data or earlier evaluation
results can predict, before an evaluation is run, how much the standard deviation (SD) of the score grows when the same
budget is concentrated on fewer sources. We use the classical design effect with an intraclass correlation $\rho$
estimated on data disjoint from those used to check the prediction. On GSM-Symbolic, 100 problems drawn as 10
templates $\times$ 10 variants have an SD of 9.2\pp\ against 2.7\pp\ from 100 templates; pilot data on 15 other
questions predict the ratio as 3.49 against 3.52 measured on held-out questions (medians over 20 splits; single splits
err by $-13\%$ to $+16\%$), once the actual sampling plan is included. On SWE-bench Verified, 10 randomly selected public
submissions predict the median ratio for 119 other submissions within 3\% for 2 or more repositories, and 3 submissions
are nearly as good. Dropping the finite-population terms from the design prediction gives errors of 13--21\% on
GSM-Symbolic, where it reduces to the textbook formula, and 25--35\% on SWE-bench. Widening iid intervals by the pilot $\sqrt{\deff}$ brings pass@1 coverage on
concentrated designs to 0.93--0.98, but not pass@16 (0.85, 0.76). We also report two lessons for anyone validating such
predictions: split pilot and held-out data by question source, not by difficulty version, and do not read a good fit of
the formula as evidence of source clustering (MATH-500). The method is not new; we contribute a held-out test of its
advance predictions and a recipe for using them.
\end{abstract}

\noindent\textbf{Code:} \url{https://github.com/Guang627/deff-eval}

\section{Introduction}

Benchmarks for language models increasingly contain groups of related problems: GSM-Symbolic instantiates each
grade-school template many times \citep{mirzadeh2024gsmsymbolic}, SWE-bench draws all tasks from 12 Python repositories
\citep{jimenez2024swebench,openai2024verified}, and many reading or agentic suites attach several questions to one
passage or environment. A team with a fixed budget, say 40 SWE-bench tasks or 100 math problems, therefore makes two
choices, not one: how many problems, and how many independent sources they come from. For teams that evaluate many
checkpoints or many agent variants, as automated research pipelines do, this choice is repeated for every run, and the
precision it buys decides which differences between systems can be trusted.

That the second choice matters is known. \citet{miller2024errorbars} recommends clustered standard errors when questions
share a passage and gives a power analysis with a cluster-adjusted sample-size formula. Survey sampling quantifies the
variance cost of cluster sampling by the design effect $1+(m-1)\rho$ for $m$ units per cluster
\citep{kish1965survey}. What is missing is evidence on how accurate such \emph{advance} predictions are for real LLM
evaluation results, and on which details of the sampling plan they must include. Miller's planning example is analytic
and is not checked against observed spread.

We test the predictions directly. On two source-labelled benchmarks and one composition control we estimate $\rho$ on
pilot data, predict the SD of each allocation of a fixed budget, and compare with the SD measured over repeated draws
from held-out data. On GSM-Symbolic the held-out data share no original questions with the pilot; on SWE-bench they are
submission-disjoint, with pilot and held-out submissions evaluated on the same 500 tasks. Our findings:
\begin{itemize}
\item For pass@1, the prediction matches the held-out SD ratio to within a few percent (median over splits) on
GSM-Symbolic and, for a typical submission, on SWE-bench Verified with 2 or more repositories, provided it uses the
actual sampling plan. Finite-population corrections are not a detail: dropping them from the design prediction gives errors of 13--35\%.
\item On SWE-bench, the public results of a handful of other submissions are enough to plan the evaluation of a
typical new one.
\item Widening iid intervals by the pilot $\sqrt{\deff}$ restores near-nominal pass@1 coverage, including on designs with
so few sources that a cluster bootstrap on the evaluated set falls well short (0.85 at GSM $10\times10$, 0.53 and 0.00
at SWE $2\times20$ and $1\times40$); for pass@16 it does not.
\item Two pitfalls: splitting pilot and held-out data by difficulty version leaks question sources, and the formula fits
composition effects as well as source clustering.
\end{itemize}
The estimator, the design effect and the variance formulas are standard and we make no methodological novelty claim.
Our contribution is an empirical one: a check, on real LLM evaluation results, of how accurate these advance predictions
are on held-out data, and of how much the finite-pool corrections matter.

\section{Related Work}

\paragraph{Uncertainty in LLM evaluation.} \citet{miller2024errorbars} argues for standard errors on evals, clustered
when questions are grouped, and for power analysis before running them. \citet{madaan2024variance} measure seed and
benchmark variance across models, and \citet{bowyer2025clt} show that CLT-based intervals are unreliable for evaluations
of a few hundred items or fewer. Efficient-evaluation work such as tinyBenchmarks \citep{polo2024tinybenchmarks}
reduces the number of items needed for a given accuracy; our question is complementary: given the number of items, how
the choice of sources changes precision. Per-problem models of success probabilities, such as beta-binomial fits for
pass@$k$ \citep{kazdan2025}, describe each problem's attempts and inherit whatever between-source variance the drawn
problems carry.

\paragraph{Clustered inference with few clusters.} Cluster-robust standard errors and the cluster bootstrap are biased
downward when the number of clusters is small \citep{cameron2008bootstrap,mackinnon2023guide}. Small-sample corrections
such as CR2 \citep{bell2002bias,pustejovsky2018small} and the wild cluster bootstrap
\citep{cameron2008bootstrap,mackinnon2017wild} improve matters but still rely on the clusters present in the analysed
sample. A benchmark concentrated on 2--10 sources has very few, and with one source per stratum no within-sample
method can estimate between-source variance at all. Our approach instead borrows that variance from outside the
evaluated set.

\paragraph{Survey sampling.} The two-stage variance with finite-population corrections that we use is textbook material
\citep{kish1965survey,lohr2021sampling}; we use it unchanged.

\paragraph{Planning cluster randomised trials.} Planning a study from an intra-cluster correlation estimated in a pilot
is standard practice for cluster randomised trials, and \citet{eldridge2016pilot} show by simulation that pilots are
usually too small to estimate that correlation precisely enough for the main trial's sample size. Our pilot-size
analysis (Section~\ref{sec:robust}) asks the same question for LLM benchmarks, where the setting differs in three ways:
the evaluated set often has only 2--10 sources, the source pools are small enough that finite-population corrections
dominate, and pass@$k$ for $k>1$ is a non-linear statistic of the attempts.

\section{Setup}

\paragraph{Score, target and tiers.} A benchmark draw is a set of problems; each problem receives $n$ attempts and the
score is the mean of the unbiased per-problem pass@$k$ estimator \citep{chen2021codex}. The target is fixed in advance as
the score over the whole held-out pool. We compare \emph{tiers} that hold the problem budget (and, on GSM-Symbolic, the
difficulty mix) fixed and vary only the number of sources: $R$ sources $\times$ $V$ problems per source.

\paragraph{Naive prediction.} With $\rho$ the intraclass correlation of the per-problem estimate between problems from
the same source, the SD ratio of an $R\times V$ tier to a one-per-source baseline is
\begin{equation}
  \frac{\mathrm{SD}(R\times V)}{\mathrm{SD}(\text{baseline})} \approx \sqrt{\deff} = \sqrt{1+(V-1)\rho}.
  \label{eq:naive}
\end{equation}

\paragraph{Design prediction.} Real draws differ from the textbook case: they are stratified, and sources and problems
are drawn without replacement from finite pools. For a stratified two-stage design with weights $w_\ell$, $t_\ell$ of
$T_\ell$ sources in stratum $\ell$ and $V$ of the $N_j$ problems of source $j$,
\begin{equation}
\begin{split}
  \mathrm{Var} = \sum_\ell w_\ell^2 \Bigg[ &\Big(1-\frac{t_\ell}{T_\ell}\Big)\frac{\sigma^2_{b,\ell}}{t_\ell}
  + \frac{\sigma^2_{a,\ell}}{t_\ell V} \\
  &+ \frac{1}{t_\ell V}\Big\langle\Big(1-\frac{V}{N_j}\Big)\sigma^2_{w,j}\Big\rangle_{\ell} \Bigg],
\end{split}
  \label{eq:design}
\end{equation}
where $\sigma^2_b$ is the between-source variance, $\sigma^2_{w,j}$ the within-source variance, $\langle\cdot\rangle_\ell$
an average over the eligible sources of stratum $\ell$, and $\sigma^2_a$ the attempt noise of the per-problem estimate.
The predicted ratio is the square root of \eqref{eq:design} for the tier over \eqref{eq:design} for the baseline. On
GSM-Symbolic every template has $N_j=20$ variants and the components are pooled one-way ANOVA estimates from the pilot.
On SWE-bench there is one stratum, $T$ is the number of repositories with at least $V$ tasks, $\sigma^2_b$ is the
variance of their mean resolve rates, $\sigma^2_{w,j}$ uses each repository's own size, and $\sigma^2_a=0$; we
compute the tier-to-baseline variance ratio separately for each pilot submission, average these ratios over the pilot
submissions, and take the square root. We compute the design prediction for pass@1 only.

\paragraph{Held-out protocol.} Table~\ref{tab:protocol} lists the splits. In each split the pilot estimates $\rho$ and
the components; the held-out part is used only to measure the SD of each tier over repeated draws. On GSM-Symbolic the
split is made by original GSM8K question (Section~\ref{sec:leak}), so pilot and held-out data share no questions; on
MATH-500 they share no problems. On SWE-bench the split is submission-disjoint: the pilot is a set of randomly selected
submissions, and pilot and held-out submissions are evaluated on the same 500 tasks. This mimics the realistic case of
planning the evaluation of a new system from earlier results on the same benchmark.

\begin{table}[t]
\centering\footnotesize
\setlength{\tabcolsep}{0pt}
\begin{tabular*}{0.8\textwidth}{@{\extracolsep{\fill}}lllr@{}}
\toprule
Benchmark & Pilot (est.\ $\rho$) & Held-out & Draws/Splits \\
\midrule
GSM-Symb.    & 15 orig.\ Qs            & 85 orig.\ Qs   & 1{,}000/20 \\
             & \scriptsize(15/15/10$\times$4, $n{=}16$) & \scriptsize(85/85/40 templ.) & \\
\addlinespace[4pt]
SWE-bench V. & 10 rand.\ subm.         & 119 subm.      & 300/100 \\
\addlinespace[4pt]
MATH-500     & 150 rand.\ prob.        & 350 prob.      & --/20 \\
\bottomrule
\end{tabular*}
\caption{Pilot and held-out splits. GSM-Symbolic and MATH-500 are split by question; SWE-bench is split by submission,
and both parts use the same 500 tasks.}
\label{tab:protocol}
\end{table}

\paragraph{Data.} \emph{GSM-Symbolic} \citep{mirzadeh2024gsmsymbolic}: main, P1 and P2 sets, 250 templates
derived from 100 GSM8K questions. Of the 50 released instances per template we use the first 20, so $N_j=20$ in
\eqref{eq:design} refers to this evaluation pool; we generated 64 attempts per problem with
Qwen2.5-Math-1.5B-Instruct \citep{yang2024qwen25math} using vLLM (temperature 0.6, top-p 0.95, at most 2{,}048 new
tokens) and graded them with math-verify; each evaluation subsamples $n=16$. Budget: 100 problems,
40/30/30 by level; baseline $100\times1$; target the 0.4/0.3/0.3 level-weighted pass@$k$ from all 64 attempts.
\emph{SWE-bench Verified} \citep{jimenez2024swebench,openai2024verified}: per-task resolved lists
(\texttt{results.json}) of the public submissions in the SWE-bench \texttt{experiments} repository
(\texttt{evaluation/verified}), retrieved in November 2026; 135 submissions had such a file, of which we keep the
129 with resolve rate 0.1--0.9 (the latest is dated September 2026); the list of submission identifiers is released
with our code. There are 500 tasks in 12 repositories, one attempt per task, so pass@1 only. Budget: 40
tasks; baseline 40 tasks at random from the 500. The target for each submission is its resolve rate over all 500 tasks;
the estimate is the plain mean of the 40 drawn tasks and every interval is centred on it. Repositories are drawn
uniformly among the eligible ones, so tasks of ineligible repositories (all but three at $1\times40$) have inclusion
probability zero, and the coverage we report is empirical over these submissions, not guaranteed by design. We keep the
plain mean rather than a design-weighted estimator because we study the spread of the score that the actual
task-selection rule produces; at $1\times40$ the tasks of nine repositories cannot be drawn at all, so no weighting of
that sample gives a design-unbiased estimate of the 500-task score. We report the resulting bias separately
(Section~\ref{sec:swe}).
\emph{MATH-500} \citep{lightman2023verify}: same model and sampling as GSM; no source labels, so groups are subject
$\times$ level cells. All model generations were run once, before the analyses; the robustness analyses in
Section~\ref{sec:robust} reuse the same answers.

\section{Results}

\begin{figure}[t]
\centering
\includegraphics[width=\textwidth]{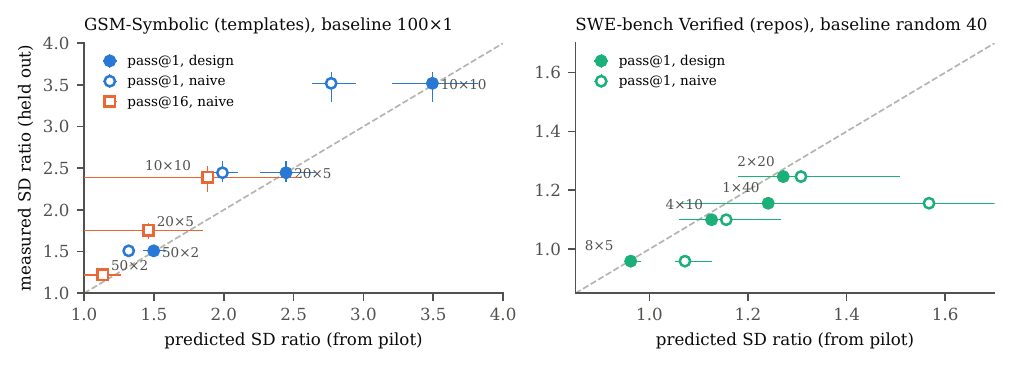}
\caption{Pilot prediction of the SD ratio (tier vs.\ baseline) against the ratio measured on held-out data. Points are
medians over splits; bars are 5--95\% ranges over splits. Filled: design prediction \eqref{eq:design}; open: naive
\eqref{eq:naive}. Left: GSM-Symbolic, 100 problems. Right: SWE-bench Verified, 40 tasks; measured value is the median
over held-out submissions.}
\label{fig:main}
\end{figure}

\begin{table}[t]
\centering\small
\setlength{\tabcolsep}{3.2pt}
\begin{tabular}{@{}llcccc@{}}
\toprule
 & tier & pilot $\rho$ & naive & design & measured \\
\midrule
GSM pass@1  & 50$\times$2  & 0.74 & 1.32 & 1.50 & 1.51 \\
            & 20$\times$5  & 0.74 & 1.99 & 2.45 & 2.45 \\
            & 10$\times$10 & 0.74 & 2.77 & 3.49 & 3.52 \\
GSM pass@16 & 50$\times$2  & 0.28 & 1.13 & -- & 1.22 \\
            & 20$\times$5  & 0.28 & 1.46 & -- & 1.76 \\
            & 10$\times$10 & 0.28 & 1.89 & -- & 2.39 \\
\midrule
SWE pass@1  & 8$\times$5   & 0.037 & 1.07 & 0.96 & 0.96 \\
            & 4$\times$10  & 0.037 & 1.16 & 1.13 & 1.10 \\
            & 2$\times$20  & 0.037 & 1.31 & 1.27 & 1.25 \\
            & 1$\times$40  & 0.037 & 1.57 & 1.24 & 1.16 \\
\midrule
MATH (cells) & A vs.\ B & 0.40 & 3.14 & -- & 2.87 \\
\bottomrule
\end{tabular}
\caption{Predicted and measured SD ratios (medians over splits). Pilot $\rho$ 90\% ranges: GSM pass@1 0.66--0.85,
pass@16 0.00--0.61; SWE 0.027--0.067.}
\label{tab:deff}
\end{table}

\subsection{GSM-Symbolic: the gap is the sampling plan}

With the same 100 problems, attempts and difficulty mix, the SD of pass@1 over draws from the full pool
(Table~\ref{tab:gsmtiers}) grows from 2.65\pp\ ($100\times1$) to 3.88, 6.62 and 9.24\pp\ as the problems are
concentrated on 50, 20 and 10 templates. Templates explain about three quarters of
the variance of per-variant success (0.74--0.84 by level), so variants of a template add little information.
Table~\ref{tab:gsmtiers} shows that the effect shrinks with $k$ but remains large at pass@16 (2.3 to 5.7\pp), and that
iid intervals keep their width while their coverage collapses.

\begin{table}[t]
\centering\small
\setlength{\tabcolsep}{3.5pt}
\begin{tabular}{@{}lccccc@{}}
\toprule
 & \multicolumn{3}{c}{SD (pp)} & \multicolumn{2}{c}{coverage, pass@1} \\
\cmidrule(lr){2-4}\cmidrule(l){5-6}
tier & pass@1 & pass@4 & pass@16 & iid & template boot. \\
\midrule
100$\times$1 & 2.65 & 2.33 & 2.31 & 0.99 & 0.98 \\
50$\times$2  & 3.88 & 3.38 & 2.81 & 0.92 & 0.97 \\
20$\times$5  & 6.62 & 5.44 & 3.93 & 0.69 & 0.93 \\
10$\times$10 & 9.24 & 7.79 & 5.67 & 0.51 & 0.85 \\
\bottomrule
\end{tabular}
\caption{GSM-Symbolic, 100 problems, $n=16$, 1{,}000 draws from the full pool (bias $<0.7$\pp\ everywhere). Mean iid
interval width is 13.4--13.8\pp\ in every tier.}
\label{tab:gsmtiers}
\end{table}

The naive rule \eqref{eq:naive} with the pilot $\rho=0.74$ underpredicts the held-out ratio by 13--21\% (ratios of the
medians in Table~\ref{tab:deff}). The reason is concrete: the $100\times1$ baseline draws 40 of 85 main and 30 of 40 P2
templates without replacement, which shrinks the baseline SD, and the textbook formula ignores this. The design
prediction \eqref{eq:design} gives 1.50, 2.45 and 3.49 against 1.51, 2.45 and 3.52 measured on held-out draws
(Figure~\ref{fig:main}, left). These measured ratios come from the held-out pools and so differ slightly from the
full-pool ratios implied by Table~\ref{tab:gsmtiers} (1.46, 2.50 and 3.49).

\subsection{SWE-bench: planning from historical submissions}
\label{sec:swe}

SWE-bench Verified illustrates the setting in which a pilot costs nothing: public per-task results of earlier systems
already exist, and a team wants to plan the evaluation of the next one. Repositories are far less homogeneous than
templates: the median share of outcome variance between repositories is 0.034 (IQR 0.024--0.049 across submissions).
Concentrating 40 tasks on fewer repositories still raises the SD, from 7.1\pp\ (random) to 9.0\pp\ ($2\times20$), and
lowers iid coverage from 0.95 to 0.83 at $1\times40$.

\paragraph{Accuracy.} With $\rho$ and the variance components from 10 randomly selected submissions, the design prediction is within
3\% of the median measured ratio over the other 119 submissions at $8\times5$, $4\times10$ and $2\times20$
(Figure~\ref{fig:main}, right); across 100 random choices of the 10 pilot submissions, the error at $2\times20$ has a
median of $+2\%$ and a 5--95\% range of $-7\%$ to $+18\%$. At $1\times40$ only django, sympy and sphinx are eligible and
40 tasks are most of sphinx (44); the naive rule overpredicts (1.57 vs.\ 1.16) and the design version is closer but
still high (1.24). The ratio below one at $8\times5$ is a finite-population effect: only 10 repositories have at least 5
tasks, so drawing 8 of them is close to stratifying by repository, and the between-repository term in \eqref{eq:design}
is multiplied by $1-8/10$, while the random baseline lets repository shares fluctuate.

\begin{figure}[t]
\centering
\includegraphics[width=0.6\textwidth]{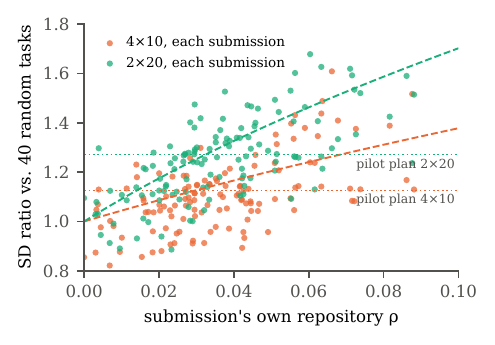}
\caption{SWE-bench Verified, 129 submissions: each submission's measured SD ratio against its own repository $\rho$
(computed on that submission's results, so this is a post-hoc check, not a held-out prediction). Dashed: naive rule
$\sqrt{1+(V-1)\rho}$. Dotted: the design prediction from 10 pilot submissions. One submission above 1.8 is outside the
axis.}
\label{fig:swe}
\end{figure}

\paragraph{A typical submission, not each one.} $\rho$ varies across submissions (5--95\% range 0.007--0.073), and so
does the measured ratio (IQR 1.13--1.37 at $2\times20$). The prediction is therefore a planning number for a typical
submission. Figure~\ref{fig:swe} shows that the spread across submissions is largely explained by each submission's own
$\rho$: at $2\times20$ the correlation between its own repository $\rho$ and its measured SD ratio is 0.82 (0.67 at
$4\times10$, but only 0.28 at $1\times40$, where three eligible repositories make the ratio erratic), and the naive rule
with the submission's own $\rho$ has a median error of $+2\%$ at $2\times20$. A team can thus plan with the pilot value
and re-estimate its own $\rho$ once its results are in. After removing task difficulty (terciles of the other submissions' solve rates) the median $\rho$ falls to 0.012, so
much of the repository effect runs through difficulty.

\paragraph{What the plan costs.} With a 40-task budget, $2\times20$ is predicted at 1.27 times the SD of 40 random tasks
(measured 1.25), so it is worth about $40/1.25^2\approx26$ random tasks, while the SD at $8\times5$ is 0.96 times that of the
random-task baseline. SD is not the whole cost: drawing repositories uniformly also biases the plain mean by $-3$ to $-4$\pp\ relative to the task-weighted target;
post-stratifying on difficulty removes most of it and lowers the SD to about 5.5\pp\ in every tier.

\subsection{Coverage: what a pilot buys after the fact}

\begin{figure}[t]
\centering
\includegraphics[width=\textwidth]{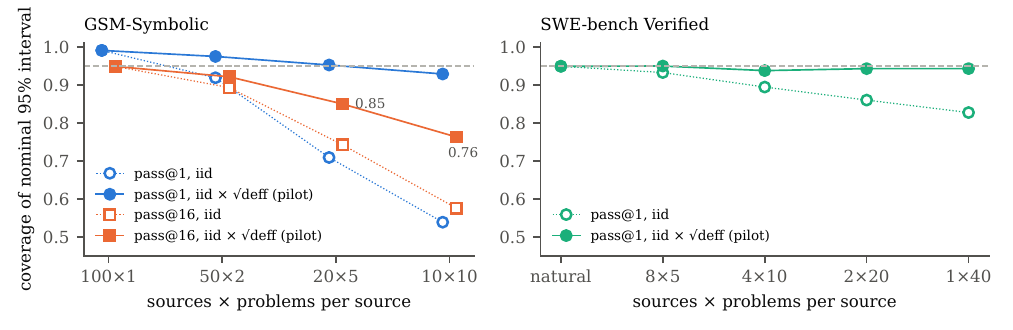}
\caption{Coverage of nominal 95\% intervals for the pool score, over held-out draws (mean over splits). Open/dotted: iid
normal interval. Filled/solid: the same interval widened by $\sqrt{1+(V-1)\hat\rho}$ with $\hat\rho$ from the pilot.
Dashed line: 0.95. SWE-bench ``natural'': the baseline of 40 random tasks.}
\label{fig:coverage}
\end{figure}

The iid interval does not see concentration: on GSM-Symbolic its width stays near 13.5\pp\ while the true spread
triples, and pass@1 coverage over held-out draws falls from 0.99 to 0.54 (Figure~\ref{fig:coverage}; the full-pool
value in Table~\ref{tab:gsmtiers} is 0.51). We widen it by the naive factor
$\sqrt{1+(V-1)\hat\rho}$, not the design ratio: the design ratio is relative to the $100\times1$ draw, whose SD is shrunk
by sampling without replacement, while the iid standard error has no finite-population correction and so matches the
infinite-pool reference the naive factor assumes. At $10\times10$ the iid standard error is about 3.4\pp; multiplied by
2.77 it gives 9.4\pp, against a measured SD of 9.2\pp. On the concentrated tiers this widening restores 0.93--0.98 on
GSM pass@1 and 0.94--0.95 on SWE-bench, including $1\times40$, where the SD prediction errs on the wide side.

The post-hoc alternative is to resample whole sources from the evaluated set, the resampling analogue of clustered
standard errors. In the same tiers (separate runs on the full pool) a template-cluster bootstrap reaches 0.97, 0.93 and
0.85 for pass@1 at $50\times2$, $20\times5$ and $10\times10$, and a repository-cluster bootstrap reaches 0.91, 0.82,
0.53 and 0.00 from $8\times5$ to $1\times40$: with one to four sources per stratum there is nothing to resample. The
pilot supplies the between-source information that a concentrated benchmark cannot supply about itself.

\subsection{MATH-500: a good fit is not a diagnosis}

MATH-500 has no source labels; its 35 subject $\times$ level cells are \emph{composition} variables. In design A each
draw takes all problems from 12 of the cells, so the mix of subjects and levels changes from draw to draw; in design B
the cell mix is fixed. Treating cells as sources, the pilot $\rho=0.40$ predicts design A's SD ratio as 3.14 against
2.87 measured, and the $\sqrt{\deff}$ correction raises coverage from 0.58 to 0.95. A reader could take this as evidence
of source clustering. It is not: fixing the composition (design B) removes the problem, with iid coverage 0.99. The
pilot also shows where the variance comes from: $\rho$ for level alone is 0.26, and for subject within level 0.18. The
formula is an accounting identity and fits whenever some grouping explains variance. To attribute inflation to shared
sources, $\rho$ must be computed for labels that still vary under the intended design, after the composition variables
are fixed.

\subsection{pass@16: where the correction is not enough}

For pass@16 the pilot $\rho$ is smaller and much noisier (0.00--0.61 across splits), the naive rule underpredicts the SD
ratio by 6--22\%, and the widened interval covers only 0.85 at $20\times5$ and 0.76 at $10\times10$. We see two possible
causes and cannot fully separate them. First, 4 variants $\times$ 16 attempts estimate the between-template component of
a non-linear statistic poorly; consistent with this, a pilot of 24 questions raises coverage to 0.88 and 0.82
(Section~\ref{sec:robust}). Second, the naive factor may miss design terms that matter more for pass@16, and we have no
design-based prediction for pass@16 that would test this. Intervals for $k>1$ therefore need a correction that has
itself been validated.

\section{Split by Question Source}
\label{sec:leak}

GSM-Symbolic's main, P1 and P2 templates are derived from the same 100 GSM8K questions: every question has a main and a
P1 template and 50 have a P2 template. Our first analysis split templates into pilot and held-out sets separately within
each difficulty level. That guarantees that no held-out \emph{template} is seen by the pilot, but not that no held-out
\emph{question} is: in the 20 splits, 28--37 of the pilot's roughly 35 original questions also reached the held-out pool
through another level. We therefore redid the split by original question (10 three-level and 5 two-level questions in
the pilot, giving the same 15/15/10 templates) and recomputed every number from the existing answers.

\begin{table}[t]
\centering\footnotesize
\setlength{\tabcolsep}{2.5pt}
\begin{tabular}{@{}lcc@{}}
\toprule
Metric & By level & By question \\
\midrule
Shared questions & 28--37 & 0 \\
Distinct questions in pilot & 32--37 & 15 \\
\addlinespace[4pt]
pass@1 $\rho$ & 0.73 & 0.74 \\
pass@1 10$\times$10 (design/meas.) & 3.48/3.52 & 3.49/3.52 \\
\addlinespace[4pt]
pass@16 $\rho$ (90\% range) & 0.42 (.09--.58) & 0.28 (.00--.61) \\
pass@16 cov.\ (20$\times$5, 10$\times$10) & 0.87/0.80 & 0.85/0.76 \\
\bottomrule
\end{tabular}
\caption{The same analysis with pilot and held-out templates split within difficulty level or by original question.}
\label{tab:leak}
\end{table}

\begin{figure}[t]
\centering
\includegraphics[width=0.55\textwidth]{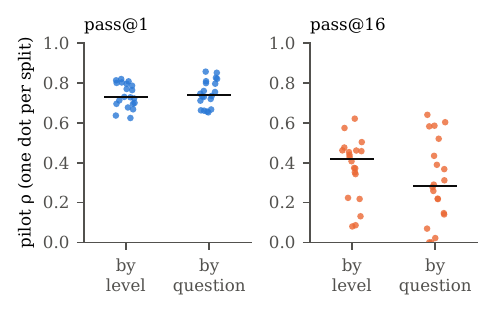}
\caption{Pilot $\rho$ in each of the 20 GSM-Symbolic splits when pilot and held-out templates are split within
difficulty level or by original question. Bars: medians.}
\label{fig:leak}
\end{figure}

Table~\ref{tab:leak} and Figure~\ref{fig:leak} show the result: pass@1 is unchanged to the second decimal, so the main finding does not rest on
the leak. pass@16 is not: its pilot $\rho$ drops from 0.42 to 0.28 and coverage drops from 0.87/0.80 to 0.85/0.76. Part
of the change need not be leakage. The level-wise pilot contained about 35 distinct questions against 15, and a
source-split pilot of 24 questions recovers a pass@16 $\rho$ of 0.42 (Table~\ref{tab:pilot}, column $\hat\rho_{16}$). Either way, the level-wise split made
the pass@16 correction look better than a clean split supports. The lesson generalises to any benchmark built by
perturbing seed items, a common construction for robustness and contamination studies: pilot and held-out data must be
split by the seed, and a pilot for $k>1$ needs enough distinct seeds.

\section{Robustness}
\label{sec:robust}

\paragraph{Spread across splits.} The medians above hide split-to-split variation. On GSM-Symbolic pass@1 the design
prediction for a single split errs by $-13\%$ to $+16\%$ (median absolute error 5\%; 83\% of split-tier pairs within
10\%). On SWE-bench the 5--95\% error range is $-1\%$ to $+3\%$ at $8\times5$, $-4\%$ to $+15\%$ at $4\times10$, $-7\%$
to $+18\%$ at $2\times20$ and $-9\%$ to $+42\%$ at $1\times40$.

\begin{table}[t]
\centering\footnotesize
\setlength{\tabcolsep}{2.5pt}
\begin{tabular}{@{}lcccc@{}}
\toprule
pilot & $\hat\rho$ & $\hat\rho_{16}$ & error [5--95\%] & coverage \\
\midrule
\multicolumn{5}{@{}l}{\emph{GSM-Symbolic, $10\times10$ (cov.\ pass@1/pass@16)}} \\
6 questions  & 0.67 & 0.19 & $-6\%$ [$-38$, $+20$] & 0.90 / 0.61 \\
9 questions  & 0.70 & 0.28 & $-5\%$ [$-26$, $+11$] & 0.91 / 0.75 \\
15 questions & 0.74 & 0.28 & $-2\%$ [$-11$, $+16$] & 0.93 / 0.76 \\
24 questions & 0.78 & 0.42 & $+4\%$ [$-12$, $+13$] & 0.93 / 0.82 \\
\midrule
\multicolumn{5}{@{}l}{\emph{SWE-bench Verified, $2\times20$}} \\
3 submissions  & 0.035 & -- & $-1\%$ [$-11$, $+13$] & 0.94 \\
5 submissions  & 0.040 & -- & $+2\%$ [$-9$, $+26$] & 0.94 \\
10 submissions & 0.037 & -- & $+2\%$ [$-7$, $+18$] & 0.94 \\
20 submissions & 0.038 & -- & $+2\%$ [$-5$, $+10$] & 0.94 \\
\bottomrule
\end{tabular}
\caption{Pilot size. $\hat\rho$, $\hat\rho_{16}$: median pilot $\rho$ for pass@1 and pass@16. Median error of the pass@1 design prediction relative to the held-out ratio, over 20 (GSM) or 100 (SWE) splits.
A larger GSM pilot leaves a smaller held-out pool, so the measured ratio also shifts slightly.}
\label{tab:pilot}
\end{table}

\paragraph{Pilot size.} We repeated the analysis with smaller and larger pilots (Table~\ref{tab:pilot}), again from the
existing answers only. On GSM-Symbolic the median error stays within $\pm6\%$ from 6 questions up, but the spread
narrows from $[-38\%, +20\%]$ with 6 questions to $-12\%$ to $+16\%$ with 15--24. Pass@1 coverage at $10\times10$ is 0.90--0.93
for every pilot size; pass@16 coverage rises with pilot size but stays below 0.95. On SWE-bench three earlier
submissions already give a median error of 1\%; 20 submissions give the narrowest range.

\begin{table}[t]
\centering\small
\setlength{\tabcolsep}{4pt}
\begin{tabular}{@{}llcccc@{}}
\toprule
 & tier & naive & no FPC & with FPC & measured \\
\midrule
GSM & 50$\times$2  & 1.32 & 1.32 & 1.50 & 1.51 \\
    & 20$\times$5  & 1.99 & 1.99 & 2.45 & 2.45 \\
    & 10$\times$10 & 2.77 & 2.77 & 3.49 & 3.52 \\
SWE & 8$\times$5   & 1.07 & 1.24 & 0.96 & 0.96 \\
    & 4$\times$10  & 1.16 & 1.41 & 1.13 & 1.10 \\
    & 2$\times$20  & 1.31 & 1.56 & 1.27 & 1.25 \\
    & 1$\times$40  & 1.57 & 1.56 & 1.24 & 1.16 \\
\bottomrule
\end{tabular}
\caption{Design prediction \eqref{eq:design} with and without the finite-population corrections (FPC), pass@1, default
pilots (medians).}
\label{tab:fpc}
\end{table}

\paragraph{When finite-population corrections matter.} Table~\ref{tab:fpc} removes the FPC terms from
\eqref{eq:design}. On GSM-Symbolic the stratified design without FPC reproduces the naive rule exactly, so the whole
13--21\% gap is the without-replacement draw of the \emph{baseline}: 40 of 85 main and 30 of 40 P2 templates. On
SWE-bench, dropping the FPC overpredicts by 25--35\% at every tier, because the tiers draw a large share of the eligible
repositories (8 of 10, 4 of 9, 2 of 7). These percentages are ratios of the medians in
Table~\ref{tab:fpc}; the medians over splits of the per-split errors are similar (12--22\% and 26--35\%). The correction matters whenever a design, tier or
baseline, draws a sizeable share of the available sources (here 35--75\% of templates per level for the GSM baseline and
29--80\% of eligible repositories for the SWE tiers), which is the usual situation for benchmarks with tens rather than
thousands of sources.

\section{A Practical Recipe}

\begin{enumerate}
\item \emph{Label sources at the level that still varies under your design}: templates, repositories, passages, or the
seed items from which perturbations were built. Fix composition variables (difficulty, subject) by stratification
first.
\item \emph{Get per-problem outcomes from outside the evaluation}: earlier systems' results on the same benchmark if they
exist (as on SWE-bench), or a pilot with several problems from each of at least 15 sources; in our GSM-Symbolic runs,
pilots of 15--24 source questions kept the 5--95\% error range within $-12\%$ to $+16\%$, and smaller pilots were much
less reliable (Table~\ref{tab:pilot}). Split pilot and evaluation data by source.
\item \emph{Predict each candidate allocation with \eqref{eq:design}}, including the finite-population terms for the
actual pools, and choose the allocation that meets the target SD. Use \eqref{eq:naive} only as a rough guide.
\item \emph{After the evaluation, report} the iid interval widened by $\sqrt{1+(V-1)\hat\rho}$ for pass@1. For pass@$k$
with $k>1$, flag the interval as uncalibrated.
\end{enumerate}
\paragraph{Worked example.} Suppose a team running a 100-problem GSM-Symbolic evaluation wants the SD of a single
model's pass@1 score to be at most 4\pp. With the pilot's components, \eqref{eq:design} predicts SDs of 2.6, 3.9, 6.4
and 9.1\pp\ for $100\times1$, $50\times2$, $20\times5$ and $10\times10$ (medians over splits), so among the four
allocations tested, $50\times2$ is the most concentrated design that meets the target; the held-out SDs (2.6, 3.9, 6.5,
9.1\pp) confirm the choice. If only 20 templates could be used,
the target would require more templates rather than more problems: with $\rho=0.74$, going from 5 to 10 problems per
template lowers the SD by only about 2\%.

\paragraph{Cost.} On GSM-Symbolic the pilot used $160\times16=2{,}560$ model calls, more than one 100-problem evaluation ($1{,}600$
calls), so a pilot run only to choose one allocation does not pay; it pays when its $\rho$ is reused across many
evaluations, as in checkpoint sweeps or automated experimentation, or when earlier results already exist. With
$\rho=0.74$, if only 10 templates can be sampled from a larger pool, the iid interval should be widened by
$\sqrt{1+9\times0.74}\approx2.8$, not by the SD ratio of 3.5, which is measured against the finite-pool $100\times1$
draw.

\section{Implications for Automated Evaluation}

Automated research and model-development pipelines evaluate many candidates against the same benchmarks, often on small
subsets to save compute, and act on the differences they see. Three consequences follow from our results. First, the
subset should be drawn to cover as many sources as possible; on GSM-Symbolic a 100-problem subset drawn from 10 templates has an SD of 9.2\pp\ against
2.7\pp\ from 100 templates, so differences of 10 points or more between candidates can be noise. Second, the variance
components needed for planning are a by-product of earlier runs: every evaluation of a previous candidate is a pilot for
the next one, so the planning cost falls to zero once a pipeline has run a few times, as on SWE-bench. Third, an
interval reported by the pipeline should carry the design effect of the subset it was computed on; an iid interval on a
concentrated subset can cover the target only about half the time (Table~\ref{tab:gsmtiers}), and a cluster bootstrap
on the same subset cannot repair this when sources are few.

\section{Limitations and Conclusion}

The two benchmarks test different kinds of transfer. GSM-Symbolic tests transfer across questions for one model with
one sampling configuration; SWE-bench tests transfer across submissions on the same tasks, for a typical submission.
Neither shows that a $\rho$ estimated for one model can be used to plan the evaluation of another: $\rho$ is a property
of a model--benchmark pair, as the spread across SWE-bench submissions shows. SWE-bench submissions are not independent
systems (many share agents or base models), and the benchmark's 12 repositories leave three for $1\times40$. The design
prediction covers pass@1 only, the robustness analyses reuse the same answers rather than new data, and MATH-500 serves
only as a negative control. Extending the design prediction to pass@$k$ is not only a matter of writing down its
variance: with $n=k=16$ the pilot gives a single 0/1 outcome per problem, which cannot separate attempt noise from
heterogeneity between variants of a template without replicate groups or further assumptions. Whether the design variance
of pass@$k$ can be predicted reliably from such a pilot is the open question that our pass@16 results leave.

We draw two conclusions at different levels of confidence. \emph{Before the evaluation}: for pass@1, a $\rho$ estimated
on disjoint pilot or earlier-evaluation data, combined with the actual sampling design, predicts the held-out SD
inflation of a concentrated benchmark to within a few percent (median over splits) on GSM-Symbolic and, for a typical
submission, on SWE-bench Verified with two or more repositories; the finite-population terms are essential. \emph{After
the evaluation}: widening iid intervals by the pilot $\sqrt{\deff}$ brings pass@1 coverage close to nominal on both
benchmarks, but pass@16 on concentrated GSM designs reaches only 0.85 and 0.76.

\bibliographystyle{plainnat}
\bibliography{references}

\end{document}